# Swift UVOT Grism Spectroscopy of Comets: A First Application to C/2007 N3 (Lulin)

D. Bodewits[1,2], G. L. Villanueva[2,3], M. J. Mumma[2], W. B. Landsman[4], J. A. Carter[5], and A. M. Read[5]

**Abstract**
We observed comet C/2007 N3 (Lulin) twice on UT 28 January 2009, using the UV grism of the Ultraviolet and Optical Telescope (UVOT) on board the Swift Gamma Ray Burst space observatory. Grism spectroscopy provides spatially resolved spectroscopy over large apertures for faint objects. We developed a novel methodology to analyze grism observations of comets, and applied a Haser comet model to extract production rates of OH, CS, NH, CN, $C_3$, $C_2$, and dust. The water production rates retrieved from two visits on this date were $6.7 \pm 0.7$ and $7.9 \pm 0.7$ x $10^{28}$ molecules s$^{-1}$, respectively. Jets were sought (but not found) in the white-light and 'OH' images reported here, suggesting that the jets reported by Knight and Schleicher (2009) are unique to CN. Based on the abundances of its carbon-bearing species, comet Lulin is 'typical' (i.e., not 'depleted') in its composition.

# 1. Introduction

Comets are relatively pristine leftovers from the early days of our Solar System. The abundance of native ices in comet nuclei is a fundamental observational constraint in cosmogony. An important unresolved question is the extent to which the composition of pre-cometary ices varied with distance from the young Sun. Our knowledge of native cometary composition has benefited greatly from missions to 1P/Halley (Grewing et al. 1988) and 9P/Tempel-1 (A'Hearn et al. 2005), from samples returned from Wild-2 (Brownlee et al. 2006), and from remote sensing of parent volatiles in several dozen comets (Bockelée-Morvan et al. 2004; DiSanti and Mumma 2008; Crovisier et al. 2009). Three distinct groups are identified based on their native ice composition ('organics-typical', 'organics-enriched', and 'organics-depleted'), and our fundamental objective is to build a taxonomy based on volatile composition instead of orbital dynamics (A'Hearn et al. 1995, Fink 2009, Mumma et al. 1993; 2003).

When comets approach the Sun, a gaseous coma forms from sublimating gases. This gas is not gravitationally bound to the nucleus and expands until it is dissociated or ionized by Solar ultraviolet (UV) light. Molecules that are released (sublimated) from native ices are called parent volatiles, and their subsequent dissociation products are called daughter- or granddaughter species. A comparison between recent optical and infrared studies of comet 8P/Tuttle revealed an unexpected dichotomy between daughter- and parent composition. The comet was 'typical' in daughter species as revealed by optical studies (A'Hearn et al. 1995), but depleted in several parent organic volatiles (Bonev et al. 2008, Kobayashi et al. 2009). The observed contradiction emphasizes the important open question of how the composition of radical (daughter or grand-daughter) species in the coma relates to that of parent volatiles and dust grains. The parentage, grand-parentage, and the evolution of a great number of species detected in cometary comae (e.g., CS, NH, CN, $C_2$, $C_3$ - A'Hearn et al. 1995, Feldman 2005, Fink 2009) are still unknown. Measuring the gaseous composition and morphology of the coma can therefore provide detailed insight into the complex relation between daughter species and their possible parents.

By quantifying the water and organic ice chemistry in the coma, the instruments on board Swift - especially the Ultraviolet Optical Telescope - can provide a unique window on comets. Observations of comet Tempel-1 in support of NASA's Deep Impact mission demonstrated the value of this approach. Two similar instruments (Swift-UVOT (Mason et al. 2007) and the Optical Monitor on board XMM (Schulz et al. 2006)) used the rapid cadence and broadband spectral filters of UVOT to trace the development of the gaseous and ice components of comet Tempel-1.

Here, we present a novel methodology for analyzing grism observations of comets, which allowed retrieval of absolute gas and dust production rates with Swift-UVOT for the first time. Grism spectroscopy is uniquely well suited for observing faint extended objects, as it combines high sensitivity with spatially resolved spectroscopy over large extended areas. The spectral ranges of the two UVOT grisms (together spanning 1800 – 6500 Å) encompass known cometary fluorescence bands such as OH (3060 Å, a principal photolysis product of $H_2O$), of native CO, and of many other molecular fragments (e.g., NH, CS, CN, fragment CO, $CO_2^+$, etc.).

We apply this methodology to C/2007 N3 (Lulin), and present quantitative results for production rates of various species (e.g., OH, NH CN, $C_2$, and $C_3$). Comet Lulin was discovered by Lin Chi-Sheng and Ye Quanzhi at Lulin Observatory. It moves in a retrograde orbit with a low inclination of just 1.6° from the ecliptic. The original orbit of comet Lulin featured $1/a_0$ of 0.000022 corresponding to aphelion distance ($2a_0$) of ~92,000 AU (Nakano 2009), and a Tisserand parameter of -1.365. These parameters identify comet Lulin as dynamically new (Levison 1996) and its recent apparition as its first journey to the inner solar system since its emplacement in the Oort cloud. It was discovered long before perihelion, and its expected favorable apparition enabled many observatories to plan campaigns well in advance of the apparition.

The outline of this paper is the following: in Sect. 2, we provide the details of our observations. In Sect. 3, we introduce the data reduction technique to clean and calibrate our comet observations. In Sect. 4, the model and assumptions used to derive gas and dust production rates are described. In Sect. 5, we compare our results with others reported to date, and discuss the composition of Lulin in the context of the emerging taxonomy for comets. We summarize our findings in Sect. 6.

## 2. Observations

Swift is a multi-wavelength observatory equipped for rapid follow-up of gamma-ray bursts (Gehrels et al. 2004). Its Ultraviolet and Optical Telescope (UVOT; Mason et al 2004; Roming et al 2005; Breeveld et al 2005) has a 30 cm aperture that provides a 17 x 17 arc min field of view with a spatial resolution of 0.5 arcsecs/pixel in the optical/UV band (range 1700 - 6500 Å). Seven broadband filters allow color discrimination, and two grisms provide low-resolution spectroscopy at UV (1700 – 5200 Å) and optical (2900 – 6500 Å) wavelengths. These grisms provide a resolving power ($R = \lambda/\delta\lambda$) of about 100 for point sources.

Swift/UVOT observed comet Lulin 16 times (post-perihelion) on 28 January, 16 February, and 4 March 2009 for a total of 18.8 ks. On each day, observations were taken 1 to 2 hours apart, providing a coarse sampling of cometary activity and changes with heliocentric distance. The UV grism was used for two exposures (each approximately 515 s in duration) during the first visit on 28 January; afterwards, the comet was too bright and moved too quickly for grism use. The high count rate and telemetry limitations did not allow use of the 'event-mode', in which all photons are time-tagged. The broadband images and X-ray observations (obtained in February and March) will be discussed in a separate paper (Carter et al., in preparation).

The observing geometry and other parameters are summarized in Table 1. During the grism observations, the comet was 1.24 AU from the Sun and 1.07 AU from Earth, with a total visual magnitude of $m_v \sim 7$ (Hicks et al. 2009). At this geocentric distance, the field of view of UVOT corresponds to (7.9 x 7.9) x $10^5$ km at the comet. We improved the signal to noise ratio by binning the pixels by a factor of four in each dimension[a]. Each binned pixel samples an area 1,722 km x 1,722 km at the comet, on 28 January.

In the wavelengths covered by UVOT, comets are seen in sunlight reflected by cometary dust, with several bright molecular emission bands superposed. One of the unique features of a grism detector is that it provides a large scale, 2-dimensional spatial-spectral image. Information in the cross-dispersion direction represents a spatial profile for the wavelength sampled, but the spatial and spectral information are blended in the dispersion direction. Thus the interpretation of grism observations of extended objects requires modeling to disentangle the spatial and spectral information.

---

[a] Any further reference to 'pixel' in this paper implies a binned pixel, i.e. a square of 4 x 4 UVOT pixels.

A raw grism image of comet Lulin in sky coordinates is shown in Figure 1a and the observing geometry is illustrated by the inset in the bottom right corner. The projections of the direction to the Sun and the orbital motion of the comet are approximately opposed. The comet was not tracked but was allowed to move through the FOV. During the 515 seconds exposure, the comet moved 1.8 arcseconds in right ascension and 6.4 seconds in declination, corresponding to a motion of +11 pixels in the dispersion direction, and -2 pixel in the orthogonal direction. We compensated for this motion during data reduction (Section 4).

The dispersed spectrum falls on the anti-sunward side of the comet and has an angle of about -10 degrees from the comet-Sun axis. This particular exposure was obtained with the UV grism in its nominal position. The broadband image ($0^{th}$ order) of gas and dust around the comet appear in the bottom left corner. The dispersion direction is approximately towards the top-right corner (arrow). A faint tail (arrow 2) can be seen extending in the anti-Sunward direction at a position angle of ~10 degrees with respect to the dispersion axis (the ion tail is more evident after processing, e.g. Fig 2i).

The $0^{th}$ order image approached the brightness limit of UVOT, resulting in clear saturation effects around the position of the comet's nucleus (dark spot in center of $0^{th}$ order image, under cross). Also, the interface connecting the detector buffer to Swift-UVOT's internal data processing system can only handle a limited number of events per second from the detector's buffer memory to the spacecraft memory. When photons strike the detector at a rate higher than about $2 \times 10^5$ counts/s, some data are dropped. This blanking effect can be seen southwest side of Figure 1a as a darker gradient on the CCD. Data in this area have an unknown response time and were discarded.

The $0^{th}$ order image is effectively a white filter image of the comet, and it can be used to study morphologies in the coma (section 5.6). For comparison, an image obtained with the UVW1 broadband filter (centered at 2600 Å) is shown in Figure 1b. The UVW1 image and iso-intensity contours (green) are shown in a logarithmic scale to make weak features more visible (note the comet tail).

# 3. Analysis

In order to extract the gaseous composition of the coma from UVOT grism observations, we developed a 5-step analysis routine. First, the image is cleaned by removal of fringes. Secondly, stars and electronic artifacts are removed by comparing different exposures. Thirdly, the center of the $0^{th}$ order image is determined, allowing us to establish the wavelength calibration at other positions. The fourth step uses this calibration to remove the flux contribution of the $0^{th}$ order image. In the final step we compare the residual spectrum with comet outgassing models and obtain absolute production rates. Additional details are described below and the process is illustrated in Figure 2.

### 3.1 Flat fielding
The grism images show a clear fringe pattern resulting from a correction for geometric detector distortion in the Swift pre-processing of the data. This correction method was optimized to preserve point-source photometry, but apparently introduces the fringe artifacts into images of extended objects. We created a flatfield image by starting with a dummy raw image with an exposure time of only 1 second, and transformed it to a detector image in the same way real data is transformed (Fig. 2b). A fringe pattern was removed by dividing the raw image (Fig. 2a) by the normalized flat field (Fig. 2b). The result is shown in Fig. 2c.

**3.2 Star removal**

Because comets are extended sources, background stars can be an important source of contamination. If the comet moves significantly between successive exposures, background star fields can be subtracted by comparing adjacent exposures (A and B). When Swift is re-pointed between two exposures, the distribution of stars on the CCD changes significantly, enabling us to identify and remove them. We assume that changes in the comet image are minor between the two exposures. If the pointing is similar for two subsequent exposures (as it was for comet Lulin), we first align the frames using the $0^{th}$ order comet images, and then identify stellar signatures in the shifted star fields.

The 'bad' pixels (i.e., containing stellar signatures or any other contamination) in both frames (A and B) are found by taking the absolute values of the difference between the two frames. In the |A-B| image the comet signal and steady backgrounds are cancelled, and all pixels with a value of more than $3\sigma$ larger than the median pixel value are masked. Next, we create a clean 'summed image' (C) by adding the A and B frames and filling each bad pixel with a Gaussian kernel (FWHM 10 pixels), forming an image that contains the comet (and diffuse background) but no stars. The last step is to clean the individual frames. We do this by comparing them with the clean summed image C, which gives the outlier pixels shown in the mask (Figure 2e). Filling these in with a Gaussian kernel yields a clean frame (e.g., Fig. 2f).

**3.3 Wavelength Calibration**

The wavelength anchor point is defined by the centroid of the $0^{th}$ order, and we express the wavelength as a polynomial in pixel number from the anchor point (i.e., distance) in the dispersion direction. Polynomial coefficients and flux calibrations were obtained from Swift calibration files.

The accuracy of the wavelength and flux calibration thus depends directly on the precision by which we can establish the position of the center of the $0^{th}$ order. Swift is designed for rapid pointing with a median pointing error of 3.5" (Breeveld et al 2009). We checked the pointing accuracy by comparing positions of the brighter stars in the field of view with archival photographic sky fields obtained with the UK Schmidt Telescope (available through the SAO-DSS image archive). From the 5 brightest stars in the optical, we find that the pointing was accurate to within one (binned) pixel. We assume the centroid to be centered exactly between positions of the comet at the beginning and end of the observations.

The optical magnitude of Lulin was approximately $m_v$ = 7 during our grism observations (Hicks et al. 2009), resulting in saturation of the $0^{th}$ order image near the comet's optocenter (Figure 2). This makes it very difficult to determine the center of the $0^{th}$ order image in the dispersion direction. An estimate of the orthogonal coordinate of the centroid was obtained by examining cross sectional profiles orthogonal to the dispersion direction. The coordinates found in this way agree with the predicted position of the nucleus within 1 pixel and confirm the accuracy of our method. The measured position of the image centroid (the 'x' in Fig. 2g) agrees with the position predicted for the comet midway through the observations (its predicted positions at the beginning and end of the observation are marked by black dots in Figs. 2g and 2h).

### 3.4 Coma and Background Removal

The $0^{th}$ order image is very extended and forms an important background source in our observations. It contains emission from both gas and dust. To remove it we used 'azimuthal averaging', a technique that is often used in studies of coma morphology (this is done most easily by converting the image into polar coordinates, e.g., Schleicher & Farnham 2004). An empirical profile is created from median values of annuli centered on the $0^{th}$ order centroid. This profile is the sum of local values for the sky background (which is constant across the CCD) and the $0^{th}$ order image (its intensity decreases approximately as $1/r$, where r is the distance from the position of the comet nucleus). We excluded a box 31 pixels in height centered on the y-position of the $0^{th}$ order, to avoid including the higher intensity regions of the dispersed cometary spectrum.

The coma profile is shown in Figure 3. Here, we show the resulting spectrum extracted for a (narrow) region 31 pixels in height (i.e. perpendicular to the dispersion axis) and centered on the dispersion axis. At short wavelengths, the background is dominated by the $0^{th}$ order profile (green). Its limit at longer wavelengths provides a measure of the sky background, which is about 0.44 counts $s^{-1}$ $pixel^{-1}$ and agrees well with reported typical background rates[a] (0.24 – 0.48 counts $s^{-1}$ $pixel^{-1}$, corrected for our binning).

The contributions by the $0^{th}$ order image and background emission are removed by subtracting an image constructed from the profile described above. The resulting $0^{th}$ order and residual images are shown in Figures 2h and 2i.

### 3.5 Uncertainties

The results are subject to several possible systematic uncertainties. The relative wavelength calibration is accurate within 15 Å (Kuin et al. 2009) but the absolute accuracy depends heavily on the position of the centroid of the $0^{th}$ order. Secondly, the $0^{th}$ order (which contains both UV and visible light) is often saturated and elongated (as can be seen from the shape of the $0^{th}$ order images of the stars within the field of view). Thirdly, the instrument might introduce a zero point shift of about 1 binned pixel (~14 Å). The uncertainty associated with the wavelength calibration may also influence the flux calibration, as the effective area is wavelength-dependent. A shift in the position of the $0^{th}$ order of ±2 pixels would result in a wavelength shift of about ± 28 Å at 3000 Å, which translates to a wavelength dependent uncertainty of the order of 5% in the effective area. The instrument flux calibration itself is accurate to 25% (Kuin et al. 2009). Added in quadrature, these uncertainties lead to an absolute systematic uncertainty of 25% in the measured intensity. The relative uncertainties are smaller.

---

[a] http://heasarc.nasa.gov/docs/swift/analysis/uvot_ugrism.html

As for most observational studies, uncertainties in the models used to derive production rates (e.g., uncertainties in the g-factors, velocities, lifetimes) can affect the results in systematic ways. However, these uncertainties are generally poorly understood (else they could be incorporated) and they are not statistically distributed. They are not incorporated in our analysis and will be discussed only qualitatively.

Statistical errors are dominated by stochastic effects of the (combined) large background and 0[th] order contributions. The background and comet signals have similar count rates at the center of the chip where the comet is brightest, but the background dominates the comet at the edge of the combined image (Figure 3). The resulting statistical error, given by the sum of the squares of the Poisson errors of the original data and the subtracted background image, is typically ~10% ($\pm 3\sigma$) of the comet's emission. Individual uncertainties are shown for each point in the residual spectrum (Figure 3).

# 4. Gas and Dust Production

To derive gas production rates we first produce a model image based on the estimated number density and emissivity of gas molecules and dust, convolve this with the instrument properties (i.e. spectral and spatial resolution) and compare the resulting model image with the observed grism image.

In its most developed form, the density distribution of different cometary species and their dissociation products is calculated using a vectorial model (Festou 1981) that replaced an earlier version called the Haser model (Haser 1957). The Haser model relates the abundance of parent and daughter fragments according to their respective scale lengths, through a mathematical relation. Because it involves only two parameters, it is much simpler to apply and is adequate for our analysis.

The approach begins by assuming that a spherically symmetric distribution of parent volatiles flows outward at constant speed until it is destroyed by photoionization and/or photodissociation reactions with the solar UV radiation field. The comet's distance to the Sun determines photodestruction and photoionization rates, which vary greatly among gaseous species (see Table 2). A numerical integration yields column densities as a function of distance to the projected center of the comet. In our application, we assume parent outflow velocities of $v_{gas} = 1$ km s$^{-1}$. For OH we derived the 'Haser equivalent expansion velocity' (Combi et al. 2004) of 1.4 km s$^{-1}$, and we assumed the same outflow velocities for the other daughter species. This transformation yields a more realistic spatial profile by relating a vectorial model to a set of Haser scale lengths (Combi et al. 2004). We further assumed that destruction lifetimes scale as $r_h^2$ ($r_h$ is the comet's heliocentric distance during the observations)

If the coma is optically thin and we neglect prompt emission, the number of photons emitted is the product of these column densities and the relevant fluorescence efficiencies, summarized in Table 3. All fluorescence efficiencies are scaled by $r_h^{-2}$, and for OH, CN and NH we took the dependence on heliocentric radial velocity ($v_h$; the Swings effect) into account (Schleicher 1983; Schleicher & A'Hearn 1988; Kim et al. 1989). For OH, we treated the $0-0$, $1-0$ and $1-1$ bands separately. We took the variation with heliocentric distance into account for the CN $0-0$ band, and kept the g-factor of the CN $1-0$ band at 8% that of the $0-0$ band (Schleicher 1983).

Several C$_2$ bands are sampled in our spectrum (cf., Feldman et al. 2004). The strongest are the Swan bands of the triplet system ($d^3\Pi_g - a^3\Pi_u$) between 4500 – 6000 Å. At shorter wavelengths (around 2300 Å) the Mulliken bands ($d^1\Sigma_u - X^1\Sigma_g$) arise from excitation from the singlet ground state. For the Swan bands with $\Delta v = 0$, we assumed $g_{band}$ (at 1.24 AU) = 0.071 photons $s^{-1}$ molec$^{-1}$ (A'Hearn 1982), and used ratios of 0.14, 0.49, 0.56 and 0.06 to weight the other Swan bands with $\Delta v = -2, -1, +1$ and $+2$, respectively (A'Hearn 1978). The Swan bands with $\Delta v = -1, 0, +1, +2$ were included in the model; the $\Delta v = -2$ band lies outside the pass band of the UV grism and was therefore excluded. To model the shape of the band structure we used high-resolution observations of Korsun & Lipatov (1993). For the Mulliken $\Delta v = 0$ transition we assumed a Mulliken to Swan ratio of $4 \times 10^{-3}$ following A'Hearn & Feldman (1980). For the $C_3$ band we used fluorescence efficiencies from Cochran et al. (1992) and a spectral shape based on the observations of Korsun & Lipatov (1993).

To model the dust distribution we assumed a spherically symmetric dust coma with an outflow velocity of $v_{dust} = 0.58 r_h^{-0.5}$ km s$^{-1}$ (Delsemme 1982). We then used solar UV spectra from the SOLar Stellar Irradiance Comparison Experiment (SOLSTICE, McClintock et al. 2005) on board the Solar Radiation and Climate Experiment (SORCE) to simulate the continuum contribution due to the reflection of sun light by dust and ice particles. Daily solar spectra are available on SORCE's interactive data website[a], and we used a solar spectrum obtained simultaneously with our observations. The spectrum was reddened by 10% per 100 Å (e.g., Remillard & Jewitt 1985).

The 2-D photon flux distribution for an emission line is then given by:

$$F(x,y) = \frac{g(r_h, v_h) \cdot N_{col}(x,y)}{4\pi \cdot \Delta^2}$$

where $g(r_h, v_h)$ is the fluorescence efficiency (g-factor) of the transition at heliocentric distance/velocity $r_h$ and $v_h$, $N_{col}(x,y)$ the column density of the species, and $\Delta$ is the heliocentric distance. Using this formalism, we created artificial images of the 2-D flux distribution for six different molecules (each with one or more emission features) and one image for the continuum. A monochromatic flux image $F(x,y)$ is converted into image in units of counts $C(x,y)$ by the relation:

$$C(x,y) = F(x,y) \cdot A_{eff}(\lambda) \cdot \Omega \cdot \Delta t$$

where $A_{eff}$ (l) is the wavelength dependent effective area of the instrument, W is the acceptance angle of one (binned) UVOT pixel (W = $9.4 \times 10^{-11}$ sterad), and $\Delta t$ is the total exposure time. To obtain a model image of the continuum, we convolved the solar spectrum with the 2-D dust model image.

For the Lulin observations, the apparent motion resulted in smearing of the spectral features. To account for the comet's motion during the observation we calculated its position on the CCD at small time intervals. A model image was created for each of those intervals and these were subsequently added. The separate images are all convolved with the Line Spread Function, for which we assume a Gaussian with FWHM of 3 pixels. The results are shown in Figure 4. The left panel (4a) shows the modeled static grism image with labels indicating the most prominent molecular emission features (that appear as condensations, as well as the continuum emission from the dust (which appears as a band). Panel 4b shows the comet model corrected for motion (+11 pixels in the dispersion direction and -2 pixels in the orthogonal direction). Panel 4c shows the observed grism image, which is similar to Figure 2i, except that the left side of the image (Fig. 2i) is cropped at the position of the 0[th] order, and the parts of the chip outside the effective area calibration are set to zero (and hence appear black in the image). Panel 4d shows an enhanced (2 times) image of the difference between fit (Fig. 4c) and observed (Fig. 2i) image. Most cometary emission has disappeared, and the 0[th] order image of the tail can be seen, as well as some remnant background stars.

---

[a] http://lasp.colorado.edu/sorce/data/

   To obtain absolute gas production rates, we used a least-squares technique to fit our 2D model image to the grism observations. All species were fit simultaneously by weighing their respective model images with their derived abundance.

   To estimate the comet's dust production we derived the quantity $Af\rho$ (the product of albedo, filling factor of grains in the aperture, and aperture radius $\rho$, cf., A'Hearn et al. 1984). $Af\rho$ (in units of cm) can be obtained from the observations by the ratio between the fluxes of comet and Sun at a given wavelength:

$$Af\rho = \frac{F_{com}}{F_{sun}(1AU)} \cdot \frac{4\Delta \cdot r_h^{\,2}}{\theta}$$

where $\theta$ is the size aperture equivalent to extraction area, in radians.

# 5. Results and Discussion

### 5.1 Spectral results

The results of our spectral analysis are summarized in Table 3. Figure 5 shows the cometary spectrum (circles) extracted from a rectangular region (430 x 37) x $10^3$ km wide (250 x 21 pixels along and perpendicular to the dispersion direction, respectively), along with the best fit modeled spectrum (green), and the individual spectral contributions of OH, dust (continuum), $C_2$, $C_3$, CN, CS and NH.

   To further demonstrate the significance of the different elements of our fit, we show intermediate steps of the analysis in Figure 6. Panels 6a-g show results of intermediate models that successively included OH, dust, $C_2$, $C_3$, CN, CS and NH emission features. In each successive model, we added one additional emission feature and again solved simultaneously for the abundance ratios of all included species. The best solution for that model reduced the residual emission, improved the chi-squared, and thus increased the overall quality of the fit.

   Figure 6a shows the contribution of just OH and the continuum, and illustrates very well the extent of emission features resulting from the convolution of spatial and spectral dimensions. For example, owing largely to the extended coma size, the convolved OH feature spans the entire width (251 pixels) of the detector. The effective spectral resolution does not permit separation of the three different vibrational bands of the OH A-X transition (Table 2), although the 1 − 0 transition at 2811 Å can be seen as a small bump on the OH shoulder. The addition of $C_2$ decreases the residual at the red end of the spectrum (Fig. 5b), while $C_3$ and CN remove the excess between 3500 and 4000 Å. The technique is not sensitive to very faint emission features. The NH emission at 3365 Å is statistically significant, but the putative CS emission feature around 2267 Å is only marginally significant.

Our best-fit model (Fig. 6f) is shown again, expanded and with labels, as Fig. 5. Our optimum spectral model (green curve) reproduces the observed spectrum over most of the range and results in a reduced chi-square of 1.5. However, significant structure remains in the residual spectrum, even after all modeled emissions are removed (Fig. 6f). The sharp (negative) residuals seen near 3100 Å, 3900 Å, and 4700 Å likely arise from model inadequacies for OH, CN, and $C_2$ in the inner coma, where the Haser model overestimates the number of daughter fragments. A vectorial model would likely fit the data better, but is beyond the scope of this paper. Alternatively, given the presence of two major jets and a long rotation period (Knight & Schleicher 2009), it is possible that there is significant rotational variability in outgassing, affecting the radial profiles. Note that if this were the case, a gas distribution based on the Vectorial model would not be a better representation of the observations than the Haser model used here.

There are two major regions that the model fails to fit within error, i.e., near 3000 Å, and above 4500 Å (Fig. 5). The residual spectrum shown in Figure 6f shows that our model predicts too little emission below 3000 Å, and too much just above 3000 Å. The emission mechanism of OH is well understood and it is therefore not likely that uncertainties in the ratios among the three OH fluorescence bands are responsible for the difference between model and observation. The sharp dip in the residual spectrum (Fig. 6f) suggests that our assumed OH distribution is too sharply peaked. Combi et al. (2004) demonstrate that compared with more sophisticated models, the Haser model (as used here) indeed results in a daughter species distribution that is more sharply peaked near the nucleus. Also, part of the excess emission might be attributed to the B $^2\Sigma_u$ – X $^2\Pi_g$ doublet band system of $CO_2^+$ at 2890 Å (Festou et al. 1982). We address this possibility in section 5.4.

## 5.2 Water Production Rates

The gas production rates derived from the two grism observations are summarized in Table 4. According to our measurements, comet Lulin produced $5.8 \pm 0.7 \times 10^{28}$ OH molecules/s at 00:03 UT, increasing to $6.9 \pm 0.7 \times 10^{28}$ molecules/s at 01:32 UT. Production rates for CS, NH, CN, $C_3$ and $C_2$ were also derived, as well as $Af\rho$ values at 3650 Å. Assuming a quantum yield of 86.5% for OH from water (Combi et al. 2004), we find water production rates of $6.7 \pm 0.7 \times 10^{28}$ and $7.9 \pm 0.7 \times 10^{28}$ molecules/s for the two observations.

Being a recent apparition, few results on comet Lulin have been published. Combi et al. (2009a) observed the comet from January 20 to 30 with the Solar Wind Anisotropies (SWAN) camera on the Solar and Heliospheric Observatory (SOHO) spacecraft. Water production rates were determined from the hydrogen Ly-α brightness and distribution using sophisticated propagation models (Combi et al. 2005, 2008, 2009b). Just before and after our Swift observations, Combi et al. (2009a) measured water production rates of $7.9 \pm 2.4$ and $7.7 \pm 2.3 \times 10^{28}$ molecules/s on January 27.5 and 28.5 UT, respectively (atomic hydrogen has a very long lifetime (~ $1.4 \times 10^7$ s, Huebner et al. (1992), but the SWAN results are deconvolved using advanced modeling to obtain daily averages of the water production rates). Bonev et al. directly measured the production rate of $H_2O$ ($11.71 \pm 1.1 \times 10^{28}$ molecules/s) on 2009 Jan 31.575 UT using

NIRSPEC at W. M. Keck Observatory (B. P. Bonev – priv. comm.). About a month later (2009 February 26 UT), D. Schleicher (private communication) measured a water production rate of 5.9 x $10^{28}$ molecules/s using narrowband photometry with the Hall 1.1-m telescope at Lowell Observatory. The comet's heliocentric distance then was 1.41 AU, compared to 1.24 AU during the Swift and SOHO observations. To better compare the observations, we assumed the empirical relation $Q_{H2O} \sim r_h^{-2}$ and scaled the Schleicher data accordingly.

Our SWIFT results are in excellent agreement with the other indirect and direct measurements of the water production rate and confirm the accuracy of our technique. The water production rates are compared in Figure 7.

## 5.3 Trends with aperture size

One unique feature of a grism detector is that it provides a large scale, 2-dimensional spatial-spectral image. For a point source, the dispersed image of a delta-function emission line is just the instrumental point-spread-function (PSF) at the position of the wavelength in question. The PSF defines the minimum wavelength resolution of the spectrum. For an extended source such as a comet, the situation is more complicated because spatial and spectral information are blended in the dispersion direction. Fortunately, intensity profiles along the direction orthogonal to the dispersion axis (hereafter the spatial direction) provide information on the spatial distribution about the nucleus.

Normalized gas and dust production rates are shown as a function of extraction height in Figure 8. We extracted spectra with heights between 0.5 to 125.5 pixels (861 – 216,000 km). Although technically we could extract spectra over an even larger area, we were careful to avoid the region on the detector that was affected by the blanking (Section 3). Also, at large extraction heights, the various spectral features blend together and become hard to distinguish. The production rates of most gases as well as $Af\rho$ scatter around unity, and are not sensitive to the size of the extraction height. The exceptions are CS, $C_3$, and to a lesser extent CN. The CS detection is only marginally significant (Section 5.1). The production rates of CN and $C_3$ vary inversely with increasing height. The CN and $C_3$ bands are close in wavelength, suggesting that at larger extraction heights ($>10^4$ km) it becomes increasingly difficult to distinguish their overlapping emission features.

The variation in gas production rate seen in OH, CN and $C_2$ (compared with OH) might have a physical origin. Their production rates show a slight increase relative to dust, but the emission features of dust and gas are only weakly coupled by our fitting process. This apparent variation could be caused by variability in the comet's gas production rate with time (resulting in an inhomogeneous coma), or could result from assumptions underlying the outgassing model (lifetime of parent and daughter, outflow velocity, extended source vs. origins from the nucleus).

Photochemical lifetimes and origins of $C_2$ are poorly constrained, a problem that was recently emphasized by observations of comet 8P/Tuttle, which appeared typical in its carbon-bearing daughter molecules but depleted in carbon-bearing parent molecules (Böhnhardt et al. 2008). It has been suggested that $C_2$ is at least partially produced by evaporating dust grains, and/or may be a granddaughter species of $C_2H_2$ (see e.g., Feldman et al. 2004, Combi & Fink 1997).

### 5.4 Composition

Absolute and relative production rates of OH and the minor species CS, NH, CN, $C_3$ and $C_2$ are summarized in Table 4. For these rates, we used an extraction height of ± 10.5 pixels in the cross-dispersion direction (21 pixels in total), which includes an area large enough to provide good photon statistics, allows us to distinguish neighboring features, but avoids the blanking area of the CCD. We also compare with data obtained by Schleicher et al. (priv. comm.) on Feb 26, 2009. Those data were acquired with aperture radii ranging between 31 and 102 arcseconds, equivalent to 9,500 – 33,000 km. Our gas production rates were acquired with a comparable extraction height of 18,000 km in the direction perpendicular to the dispersion axis. Both our gas and production rates are in excellent agreement with Schleicher et al., the relative abundances we derived for the minor species are consistently lower than theirs.

Comets can be classified based on their native ice composition (Section 1). By comparing a large sample of comets, A'Hearn et al. (1995) found that in the daughter species, this distinction is most apparent in the ratio between $C_2$ and CN, and to a lesser extent in the abundances of the other minor species relative to OH. The detailed origins of this difference are not understood, but may be related to different formative regions (Mumma 1993, 2003). We find a ratio $^{10}\log(C_2/CN)$ of 0.44, which is well above the threshold ratio for depleted comets of -0.18 (A'Hearn et al. 1995), suggesting that comet Lulin is 'typical' in carbon chain daughter species. It is of note that our $C_3$ abundance is low compared to other observations (Schleicher et al.), but its relative abundance $^{10}\log(C_3/OH)$ of -3.7 lies well within the range of comets of typical composition, and above that of depleted comets. (A'Hearn et al. 1995). All other abundances and the dust production rates are in good agreement with production rates of long period comets that are not depleted in their carbon chain molecules.

As discussed in section 5.1, we find an excess flux around 2900 Å that might be attributed in part to the $CO_2^+$ B – X doublet (2884 – 2896 Å). This excess flux was 2.2 and 3.3 ph $cm^{-2}$ $s^{-1}$ $Å^{-1}$ in the first and second observation, respectively. Deriving $CO_2$ production rates using a model grism image is hampered because the spatial distribution of ionic gases depends on the interaction between comet and the solar wind, and is most likely not spherically symmetric (see e.g., Wegmann et al. 1999). The lifetime of $CO_2^+$ ions in the solar UV-field is not known. We can however use the residual flux in the region 2500-3000 Å to get a crude estimate of the production rate of $CO_2$ in comet Lulin.

Two processes could produce such emission; prompt emission following the ionization of $CO_2$, and fluorescent emission by $CO_2^+$. Prompt emission would map the distribution of neutral $CO_2$ but has a relatively low efficiency. Based on the ionization cross sections measured by Gustaffson et al. (1978) and the theoretical cross sections from Padial et al. (1981), we estimate that approximately 30% of $CO_2$ photoionization reactions result in B – X band. (A similar rate applies for emission in the A – X band, which would appear as a highly dispersed (thus very faint) quasi-continuum in our spectrum.) Assuming a photo-ionization rate of 6.55 x $10^{-7}$ $s^{-1}$ (Huebner et al. 1992), this would yield an emission rate of 2.2 x $10^{-7}$ $s^{-1}$ at 1 AU. $CO_2^+$ fluorescent emission requires that the $CO_2$ molecules be ionized first, and then excited by solar UV light. Feldman et al. (1986) used a fluorescent rate of 2 x $10^{-3}$ $s^{-1}$ at 1 AU, while we find a slightly larger g-factor of 4 x $10^{-3}$ $s^{-1}$ based upon lifetimes measured by Larsson et al. (1985). Using these emission rates, we find that if the dominant process were prompt emission there would be $1.1 \times 10^{35}$ $CO_2$ molecules present within our field of view. Assuming a total $CO_2$ photodestruction rate of $2 \times 10^{-6}$ $s^{-1}$ (at 1 AU; Huebner et al. 1992) this would yield unrealistically high $CO_2$ production rates of $1.4 - 2.1 \times 10^{29}$ molec $s^{-1}$.

Emission through fluorescence requires $N_i = 6 \times 10^{30}$ $CO_2^+$ ions present within the field of view. To determine the observed tailward flux $\Phi_i$ we adopt the simple model used by Schultz et al. (1993) and Feldman et al. (1986). In this model, the ions are confined to a cylinder of radius $R < R_{FOV}$, and a length L of the tail is within the field of view (L = $4.7 \times 10^{10}$ cm). The ions are assumed to be distributed with a uniform density, and move with an average anti-sunward velocity $v_i$. The ion flux, which is equal to the $CO_2^+$ production rate, is then given by:

$$\Phi_i = \frac{N_i * v_i}{L}$$

By assuming an average ion velocity of 30 km $s^{-1}$ as obtained by measuring the Doppler shift of $H_2O^+$ ions observed in comet Halley (Schultz et al. 1993) we find a flux of $4 \times 10^{26}$ ions $s^{-1}$. From the photodestruction rates of Huebner et al. (1992) we estimate that $Q(CO_2)/Q(CO_2^+) = 0.34$. If the emission is dominated by fluorescence, the $CO_2$ production rate would be approximately $1 - 1.5 \times 10^{27}$ molec $s^{-1}$. Our observations therefore suggest an abundance relative to OH of approximately 2%, which is in excellent agreement with AKARI observations by Ootsubo et al. (2010), who found a $CO_2$ production rate of $(3.4 \pm 0.1) \times 10^{27}$ and a relative abundance $Q(H_2O)/Q(CO_2) = 4.5\%$ when the comet was 1.7 AU from the Sun.

### 5.5 Temporal variation

Two grism observations (1600 – 5100 Å) were obtained within an interval of 1.75 hours, but photon counts were on average 15% higher during the $2^{nd}$ observation. The derived production rates (Table 4) suggest that the water production increased by as much as 18% (12σ), but that the dust production decreased over this period of time. The relative abundance of minor species shows no significant differences for the two observations. Assuming radial outflow velocities of 1.0 km/s (gas) and 0.44 km/s (dust), respectively (Section 4), gas and dust would have moved about 5,340 km and 2,400 km between the two observations. Our spatial data (Figure 8) do not suggest an increased production rate in the inner 2,400 – 5,340 km of the coma. The increase in gas and dust production rates is therefore more gradual, and perhaps masked by the spatial convolution along the dispersion axis.

As noted in section 5.3, the number of daughter products at distance of $10^5$ km from the nucleus, suggests larger gas production at the time they were produced compared with later production. Along the same line of reasoning, a brief outburst 20-27h before the first observation, increasing the gas production rate by approximately 25% could explain the observed gas production profile. The observations by Bonev et al. and Combi et al. (2009, see Figure 7) suggest that gas production rates varied on a day-to-day basis by as much as 50%. Also, Knight & Schleicher (2009) report a tentative rotation period of $42 \pm 0.5$ h based on the mapping of CN jets. These variations and time-scale are in agreement with the variability we observe, and suggest a strong diurnal effect on the comet.

### 5.6 Tail and Jets

A clear tail feature appears in Figs. 1 and 2. The tail is narrow (~50,000 km) and stretches across the width of the image. Comet Lulin's retrograde orbit lies nearly in the ecliptic plane, causing the (hypothetical) ion and dust tails to appear superimposed along the orbital track, and we are not able to identify the explicit nature (ion or dust) of this tail. Amateur optical amateur observations obtained with other instruments suggest the presence of both tails a few days after our Swift observations[b].

Comparing Figs 1a and 1b, it appears that the comet tail is more evident in the $0^{th}$ order grism image (that encompasses UV and optical wavelengths) than in the UVW1 band. The UVW1 filter has its maximum sensitivity at 2600 Å with a FWHM of approximately 700 Å. Detection of an ion tail in the UVW1 filter is not likely. This filter has a long red leak, but is not very sensitive to the emission usually associated with the blue $CO^+$ comet tail system (4000 – 5000 Å). Ionic emission from $CO_2^+$ has been observed at 2890 Å (Festou et al. 1982), and our analysis indeed shows the likely presence of emission from this band (sec 5.4). More likely, dust or ice could well explain the observed tail. Continuum emission (dust, Fig. 5) is bright in the optical and would well explain the feature in the grism image. It is faint in the near UV, but could be visible through the red leak in the UVW1 filter. A dust tail would indeed be brighter in the $0^{th}$ order grism image than in the UVW1 filter.

Knight and Schleicher (2009) observed CN jets, which they used to derive the rotation period of the nucleus. The $0^{th}$ order image is essentially a white filter image of the comet and can be used to study coma features. We searched the coma-subtracted $0^{th}$ order images for jet structures but found none. We also looked for coma features in the OH emission dominated UVW1 image (Fig. 1b) by subtracting the azimuthally averaged coma profile, but could not find any evidence of jets in those images either. The jets reported by Knight and Schleicher (2009) are thus likely highly specific for the CN distribution. Further discussion of jets is deferred to a future publication.

# 6. Conclusions

The sensitivity and spatial grasp of the Swift-UVOT grisms offers a unique capability to measure the water production and mixing ratios of various fragment molecules from space, such as CS, NH, CN, $C_2$, $C_3$, and tentatively, $CO_2^+$. We developed a novel analytical procedure to reduce comet grism observations, and based on modeling of the coma, to derive absolute production rates of water, dust, and various minor gaseous species. We applied these routines to observations of comet C/2007 N3 (Lulin), observed by Swift UVOT for about 1 ks on Jan 29[th], 2009. At that point, the comet had reached the maximum optical brightness permitted for observations with the UV grism on board Swift.

The main findings of this work are as follows:

---

[b] http://apod.nasa.gov/apod/ap090207.html

- Between 00:03 and 01:32 UT, the water production rate increased from $6.7 \pm 0.7$ x $10^{28}$ to $7.9 \pm 0.7$ x $10^{28}$ molecules s$^{-1}$. Based on spatial profiles, there are no indications for an impulsive event during our observations. In accordance to observations by other teams, Lulin's gas production appears to vary by at most 25–50% on time scales of order days, which we attribute to strong diurnal variations.
- Based on our measurements, the daughter species composition of comet Lulin is typical within the larger body of long period Oort cloud comets. The comet is not depleted in its composition of carbonaceous minor species.

Our observations of comet Lulin indicate that Swift can add significantly to comet studies. Its grisms are very well suited for observing faint comets because of the large spatial grasp (as opposed to slit-based instruments) and high sensitivity. Faint, moderately active comets are not easily observed from Earth, and are relatively undersampled in current comet surveys. Swift is uniquely suited to fill in this gap.

**Acknowledgments**
We thank the Swift team for use of Director's Discretionary Time to observe comet Lulin, and for the careful and successful planning of these observations. The authors would like to thank J. Morgenthaler for helpful discussions on grism spectroscopy, and T. Farnham for valuable suggestions on the comet tail dynamics. We are grateful for the unpublished gas production rates provided by D. Schleicher and B.P. Bonev. The Digitized Sky Survey was produced at the Space Telescope Science Institute under US Government grant NAG W-2166. SOLSTICE is operated from the Laboratory for Atmospheric and Space Physics (LASP) at the University of Colorado in Boulder. We are grateful for the cometary ephemerides of D. K. Yeomans published at the JPL/Horizons web site. A NASA Postdoctoral Fellowship to DB, and grants from NASA's Astrobiology Institute and Planetary Astronomy Program to MJM supported this work.

**Table 1.** Observing log and geometry during the Swift UVOT observations of comet Lulin. Ephemeris data were obtained from the Horizons web site[a].

| Observation 1 | |
|---|---|
| Start (UT) | 2009-01-28 00:03:04 |
| Integration Time (s) | 514.5 |
| | |
| **Observation 2** | |
| Start (UT) | 2009-01-28 01:32:04 |
| Integration Time (sec) | 516.5 |
| | |
| Heliocentric Distance (AU) | 1.242 |
| Heliocentric velocity (km/s) | 5.86 |
| Geocentric Distance (AU) | 1.066 |
| Sun-Comet-Observer Angle (degrees) | 9.70 |
| Position Angle (degrees) | 285 |

**Table 2.** Parameters used in coma modeling.

| Species | Haser Lifetime[a] | | Ref. |
|---|---|---|---|
| | Parent (s) | Daughter (s) | |
| OH | $1.0 \times 10^5$ | $1.8 \times 10^5$ | A |
| CS | $5.0 \times 10^2$ | $1.0 \times 10^5$ | B |
| NH | $5.0 \times 10^4$ | $1.0 \times 10^5$ | C |
| CN | $2 \times 10^4$ | $2.1 \times 10^5$ | C |
| $C_3$ | $3 \times 10^3$ | $1.4 \times 10^5$ | C,D |
| $C_2$ | $2.2 \times 10^4$ | $1 \times 10^5$ | C,D |

References: A – Combi et al. (2004); B – Weaver et al. (1999); C – A'Hearn et al. (1995); D – Cochran et al. (1985)

---

[a] http://ssd.jpl.nasa.gov/horizons.cgi#top
[a] All lifetimes are for $r_h = 1$ AU and are scaled by $1/r_h^2$.

**Table 3.** UV Line list and parameters used in coma modeling and fluxes extracted from a rectangular area extending 10.5 pixels above and below the dispersion axis. Only stochastic errors are given. The systematic uncertainty is approximately 25%.

| Wavelength (Å) | Species | Transition | $g^b$ (ph s$^{-1}$ molec$^{-1}$) | Ref | Flux (ph s$^{-1}$ cm$^{-2}$ Å$^{-1}$) | |
|---|---|---|---|---|---|---|
| | | | | | 00:03 UT | 01:32 UT |
| 2667 | CS | 0-0 | $4.5 \times 10^{-4}$ | C | $0.3 \pm 0.1$ | $1.9 \pm 0.2$ |
| 2811 | OH | 1-0 | $1.9 \times 10^{-5}$ | D | $6.9 \pm 0.08$ | $8.3 \pm 0.07$ |
| 3064 | OH | 0-0 | $2.7 \times 10^{-4}$ | D | $96 \pm 1$ | $114 \pm 1$ |
| 3122 | OH | 1-1 | $3.4 \times 10^{-5}$ | D | $12 \pm 0.1$ | $15 \pm 0.1$ |
| 3365 | NH | 0-0 | $9.5 \times 10^{-3}$ | E | $22 \pm 1$ | $21 \pm 1$ |
| 3876 | CN | $\Delta v = 0$ | $5.5 \times 10^{-2}$ | F | $58 \pm 2$ | $64 \pm 2$ |
| 4216 | CN | $\Delta v = -1$ | $4.4 \times 10^{-3}$ | F | $4.6 \pm 0.1$ | $5.1 \pm 0.1$ |
| 4060 | $C_3$ | | 0.50 | G | $17 \pm 1.4$ | $14 \pm 1.4$ |
| 4375 | $C_2$ | Swan $\Delta v = +2$ | $3.8 \times 10^{-3}$ | H,B | $17 \pm 0.5$ | $19 \pm 0.5$ |
| 4700 | $C_2$ | Swan $\Delta v = +1$ | $3.4 \times 10^{-2}$ | H,B | $27 \pm 0.8$ | $30 \pm 0.8$ |
| 5100 | $C_2$ | Swan $\Delta v = 0$ | $7.4 \times 10^{-2}$ | H,B | $67 \pm 2$ | $75 \pm 2$ |
| 5570 | $C_2$ | Swan $\Delta v = -1$ | $3.7 \times 10^{-2}$ | H,B | $31 \pm 1$ | $34 \pm 1$ |
| **Continuum:** | | | | | | |
| 3650 | | | | | $0.01 \pm 0.002$ | $0.01 \pm 0.002$ |

**References**: A – A'Hearn & Feldman (1980); B – A'Hearn (1982); C – Jackson et al. (1982); D – Schleicher & A'Hearn (1988); E – Kim, A'Hearn & Cochran (1989); F - Schleicher (1983); G – Cochran (1992); H – A'Hearn 1978;

---

[b] All fluorescence efficiencies are for $r_h = 1.24$ AU and are scaled by $1/r_h^2$. The fluorescence efficiencies of OH, NH and CN are for the heliocentric velocity $v_h = 5.9$ km/s.

**Table 4.** Derived absolute and relative gas production rates for two SWIFT observations, and our best estimate for the production of $CO_2$. Only stochastic errors are given. The systematic uncertainty is approximately 25%. Absolute and relative gas and dust production rates obtained at UT 26 February 2009 by D. Schleicher (priv. comm.) are given for comparison.

| Species | Absolute Gas Production Rate (Q) ($10^{26}$ molecules s$^{-1}$) | | | Relative Abundance (Q/$Q_{OH}$) (%) | | |
| | This work | | Schleicher | This work | | Schleicher |
| | 00:03 UT | 01:32 UT | | 00:03 UT | 01:32 UT | |
|---|---|---|---|---|---|---|
| OH | $578 \pm 64$ | $689 \pm 61$ | $510 \pm 51$ | | | |
| CS | $0.6 \pm 0.4$ | $3.5 \pm 0.4$ | - | 0.1 % | 0.5 % | - |
| NH | $3.7 \pm 0.2$ | $3.5 \pm 0.2$ | $2.8 \pm 0.3$ | 0.6 % | 0.5 % | 0.5% |
| CN | $1.0 \pm 0.03$ | $1.1 \pm 0.03$ | $1.0 \pm 0.1$ | 0.2 % | 0.2 % | 0.2% |
| $C_3$ | $0.1 \pm 0.01$ | $0.1 \pm 0.01$ | $0.5 \pm 0.05$ | 0.02 % | 0.01 % | 0.08% |
| $C_2$ | $2.7 \pm 0.08$ | $3.0 \pm 0.1$ | $2.0 \pm 0.2$ | 0.5 % | 0.4 % | 0.3% |
| $CO_2$ | 10 | 15 | | 2 % | 2 % | |

| $\lambda$ Af$\rho$ (Å): | Af$\rho$ (cm) | Af$\rho$ (cm) | Af$\rho$ 5260 Å (cm) | log(Af$\rho$/ $Q_{OH}$) | log(Af$\rho$/ $Q_{OH}$) |
|---|---|---|---|---|---|
| 3650 | $3091 \pm 402$ | $2878 \pm 390$ | 2188 | -25.2 | -25.4 | -25.4 |

## Figures:

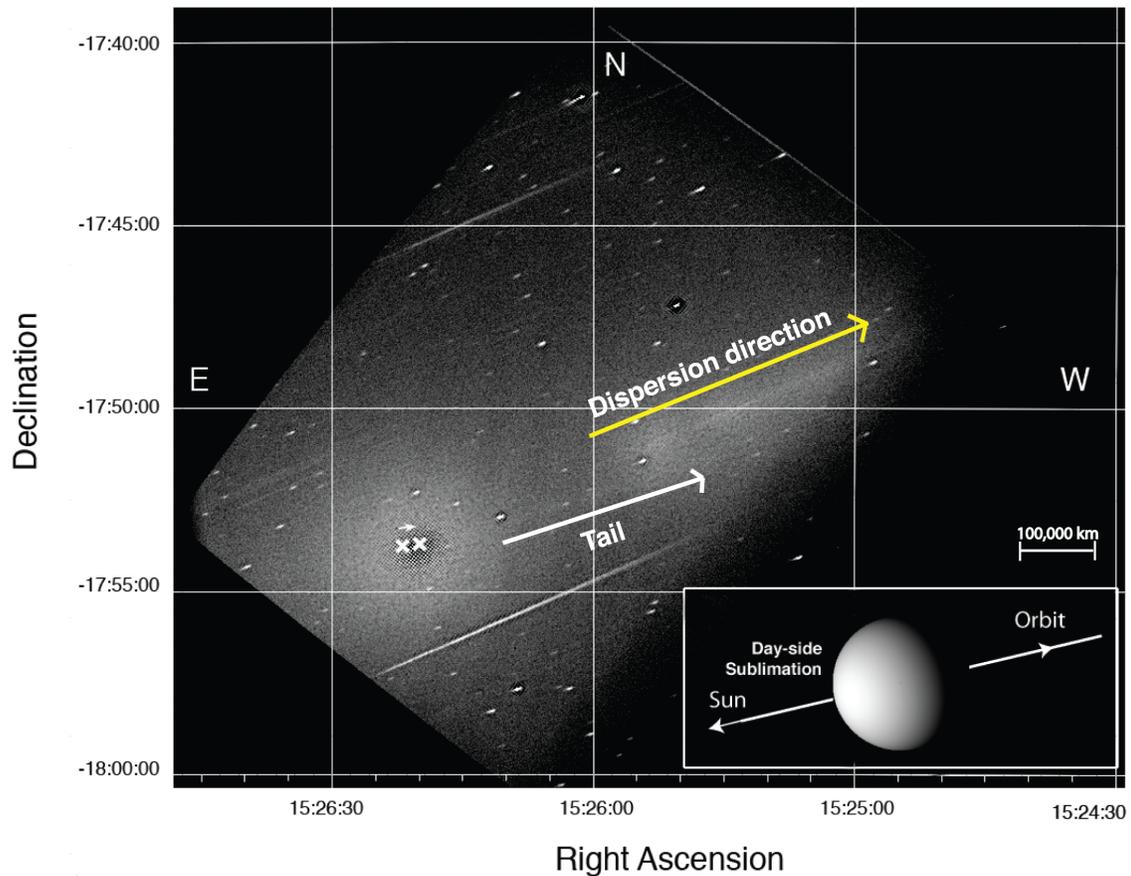

**Figure 1a.** Raw detector image and observing geometry of the first Lulin grism observation. The intensity scale is logarithmic and is stretched to enhance fainter features. The 2 crosses indicate the position of the comet nucleus at the beginning and end of the observation. The inset shows the observing geometry, along with the projected directions (arrowheads) to the sun and of the orbital motion of the comet. The dispersion axis extends to the right at an angle of ~25° with respect to the horizontal axis. The dispersed coma emission (yellow arrow) is aligned with the dispersion direction, while the tail (white arrow) appears just below the dispersion direction at an angle of ~15° with respect to the horizontal axis. The bright streaks are stellar spectra. The scale bar represents 100,000 km at the comet.

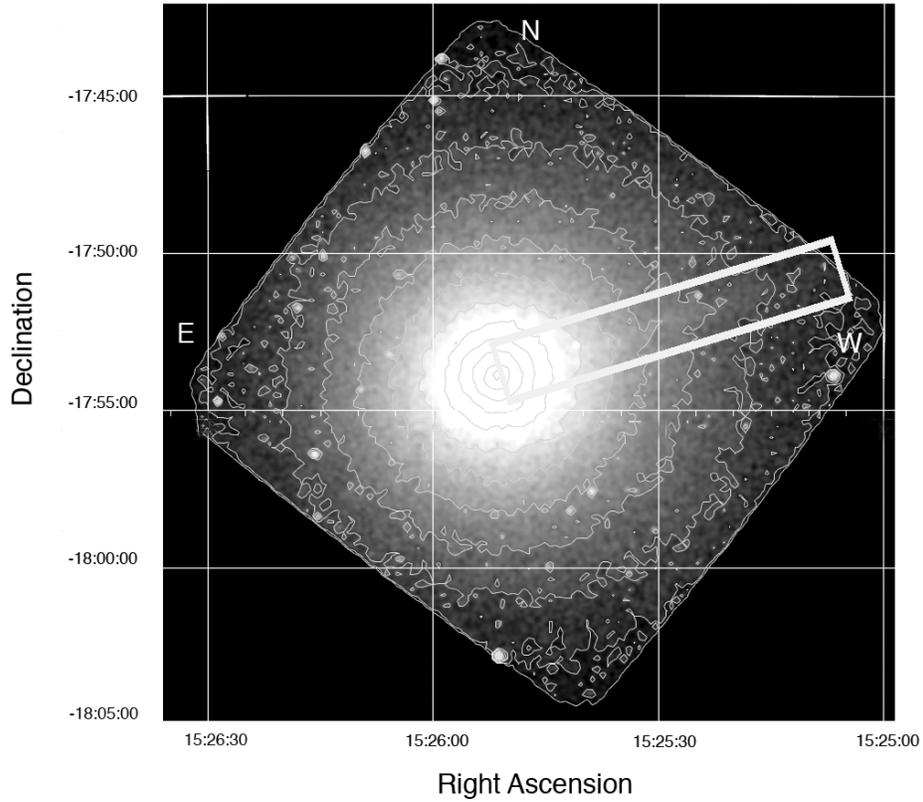

**Figure 1b.** Swift UVOT image obtained with the UVW1 filter centered at 2600 A. The iso-intensity contours are shown on a logarithmic scale to enhance fainter features. The orientation of the image is the same as in Figure 1a. The faint tail is emphasized by a grey box and can be seen extending to the right (westward).

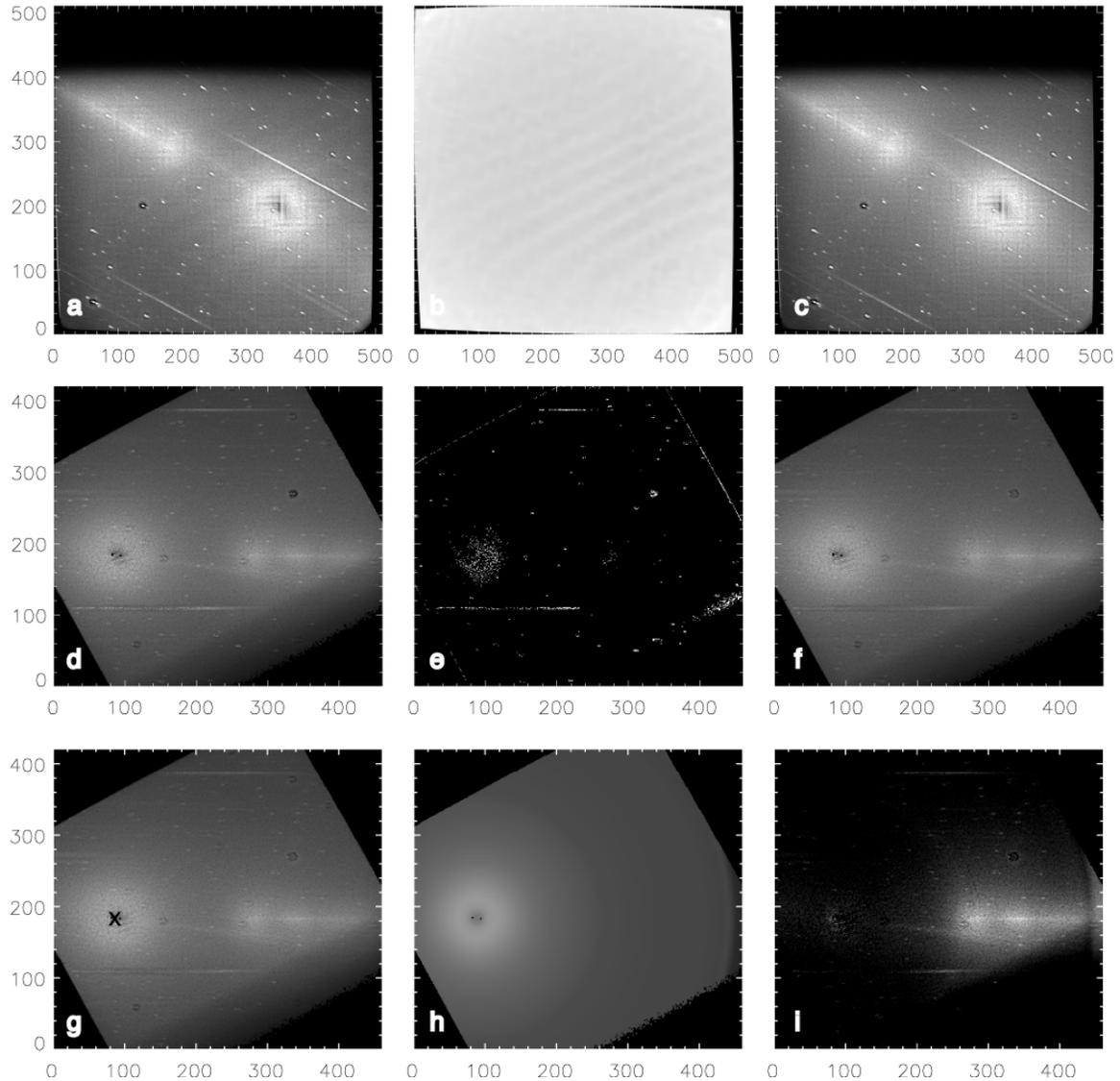

**Figure 2.** Image processing of the UVOT grism observations of comet Lulin. **a.)** Original image; **b.)** Fringe flatfield; **c.)** Fringes removed. **d.) -** as **c**, but now rotated so that the dispersion direction aligns with rows; **e.) -** Star and bad pixel mask; **f.) -** Resulting cleaned image. **g.)** – as **f**, with central position of $0^{th}$ order indicated ('x') at a time midway through the first observation. **h.) -** $0^{th}$ order image of comet and background (note saturation in the inner coma). **i.) -** Final grism image, after subtraction of $0^{th}$ order image and background signals. All images have a linear grey scale and are normalized for optimal presentation. The small black dots in **d**, **f**, **g** and **h** indicate the position of the nucleus at the beginning and end of the observations.

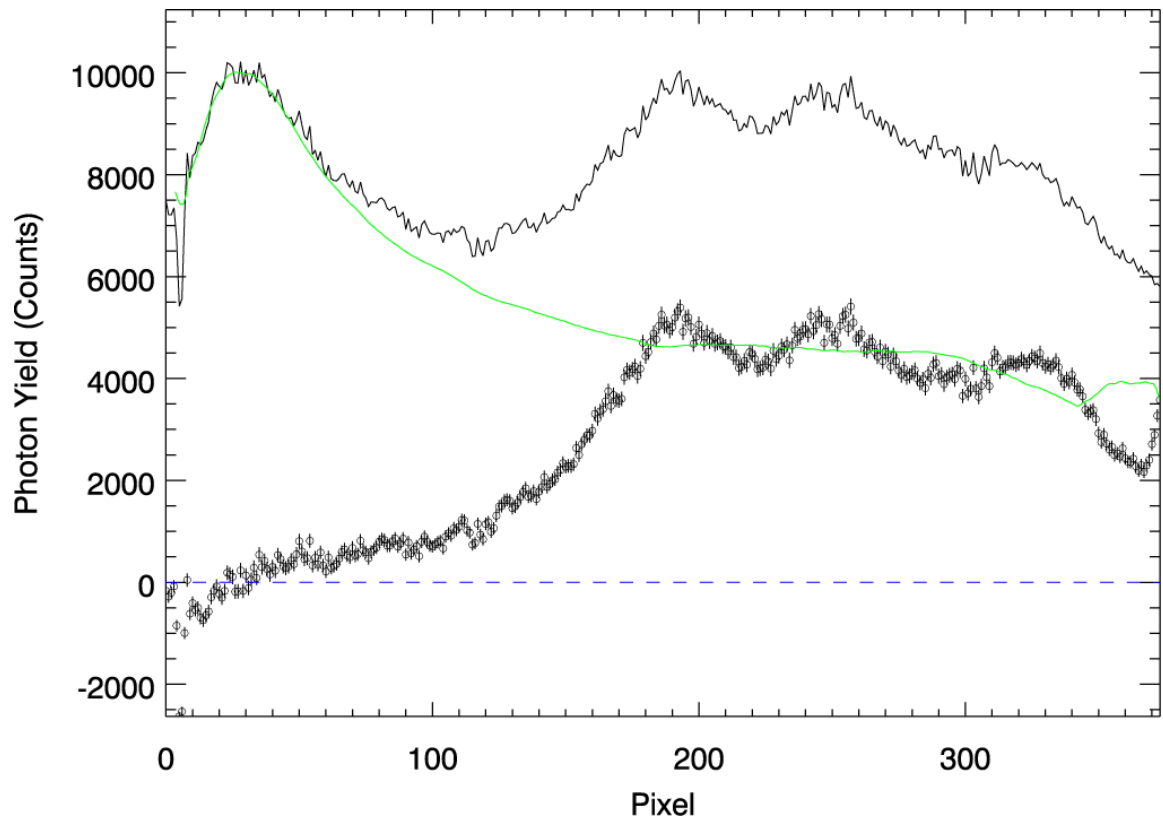

**Figure 3.** Separation of the cometary emission spectrum from the sum of background and 0<sup>th</sup> order image contributions for a region extending 21 pixels in the cross-dispersion direction.  The total emission centered on the dispersion axis is shown in black, the removed profile is shown in green, and the residuals are indicated by black circles. The statistical errors shown are ± 1σ. Multiple spectral features appear in the residual spectrum and are identified in Figures 4 and 5.

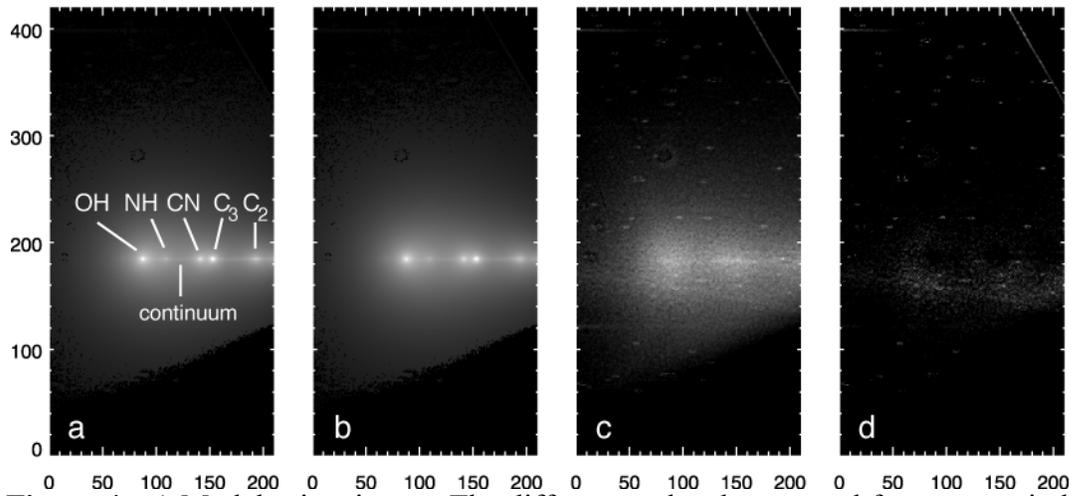

**Figure 4. a.)** Model grism image. The different molecular spectral features are indicated. The continuum can be seen as a thin bar. **b.)** similar to left panel, but smeared by comet's motion during the observation. **c.)** Observed grism image. **d.)** Residual; difference of frames **c** and **b**. The residual $0^{th}$ order image of the tail can be seen, and approximately indicates the direction to the Sun (to the left) and the comet's orbit (to the right). All panels have the same spatial scale, alignment, and intensity scale, except from **d** where the intensity was enhanced by a factor of 2. The frame dimensions (212 x 422 pixels) represent 364,434 km x 725,431 km at the comet.

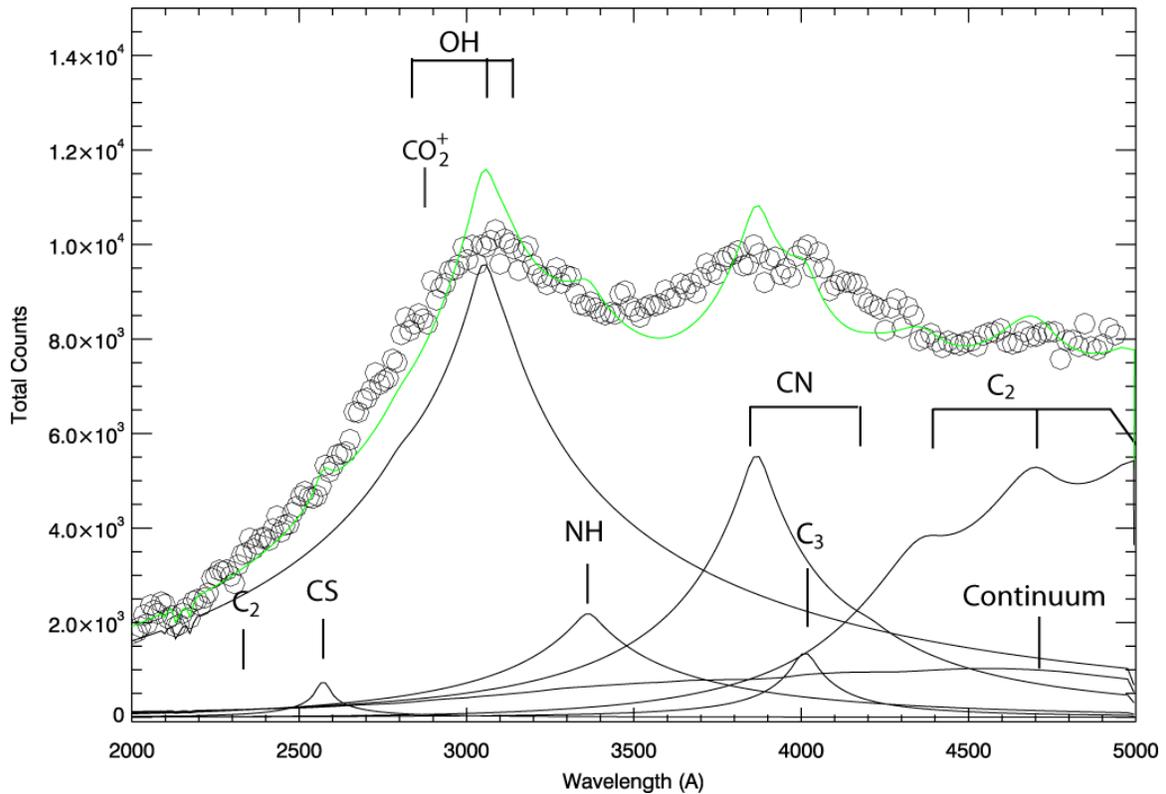

**Figure 5.** The final spectrum (extraction height 10 pixels) of C/2007 N3 (Lulin) and best-fit synthetic model. The central position of the tentative $CO_2^+$ emission is indicated. The final spectrum (circles), individual molecular bands (thin black lines), and the resulting spectral fit (green line) are shown. The statistical $1\sigma$ errors are comparable to the circles.

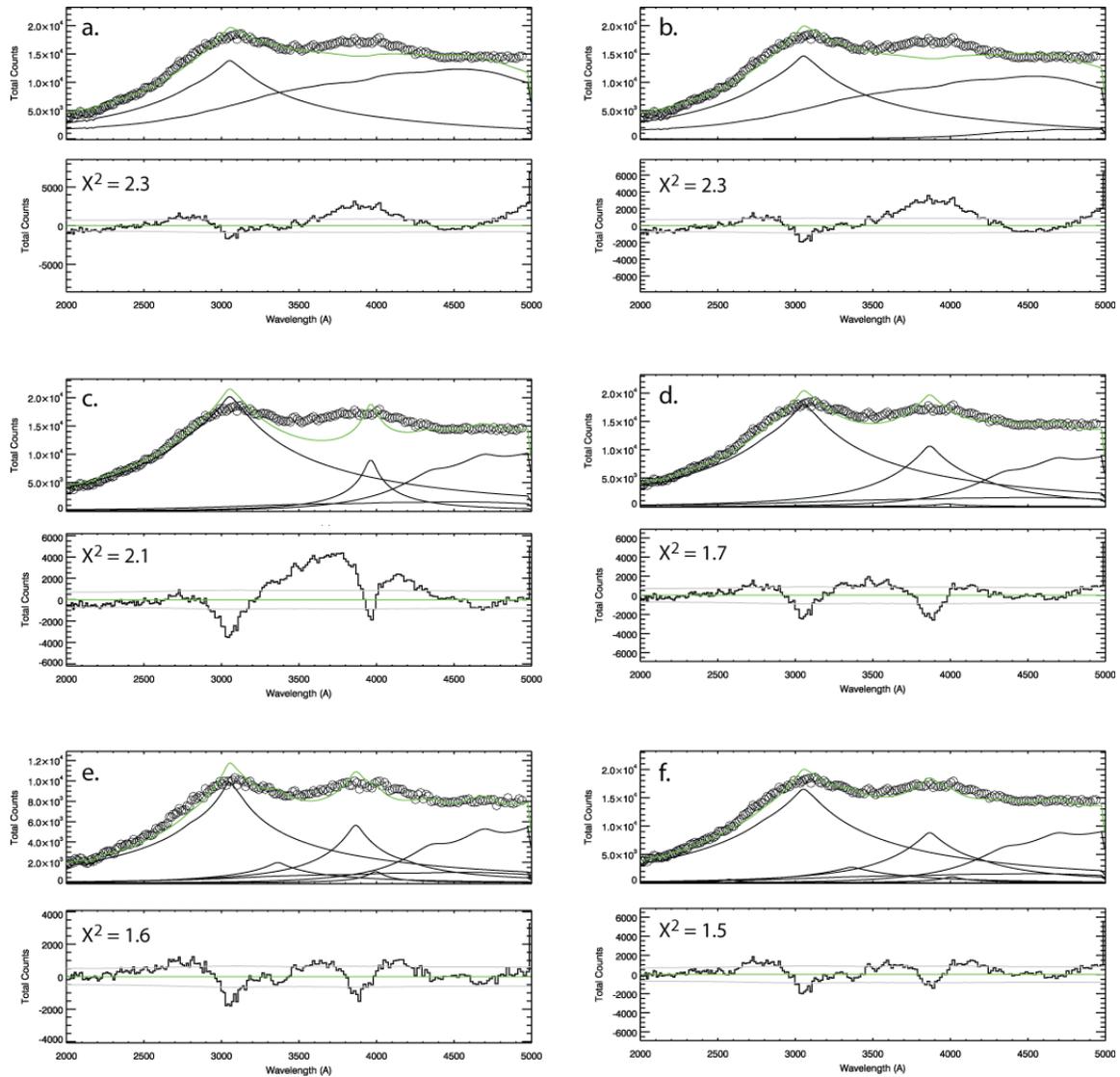

**Figure 6.** The spectrum of comet Lulin is shown (circles) as extracted from a rectangular area centered on the dispersion axis and 21 pixels in total height. In each panel (**a – f**), the green line represents the synthetic spectrum with inclusions as modeled. The residual emission after subtraction of the best-fit model is shown in the lower portion of each panel. The successive inclusion of emission features [OH and dust (panel **a**), $C_2$ (**b**), $C_3$ (**c**), CN (**d**), NH (**e**) and CS (**f**)] progressively reduces the residual emission and thus increases the overall quality of the fit. Horizontal, grey lines indicate ± 3σ errors. The sharp (negative) residuals seen near 3100 Å, 3900 Å, and 4700 Å likely arise from inadequacies of the Haser model in the inner coma, where it overestimates the intensity of daughter fragments. A vectorial model would likely fit the data better, but is beyond the scope of this paper.

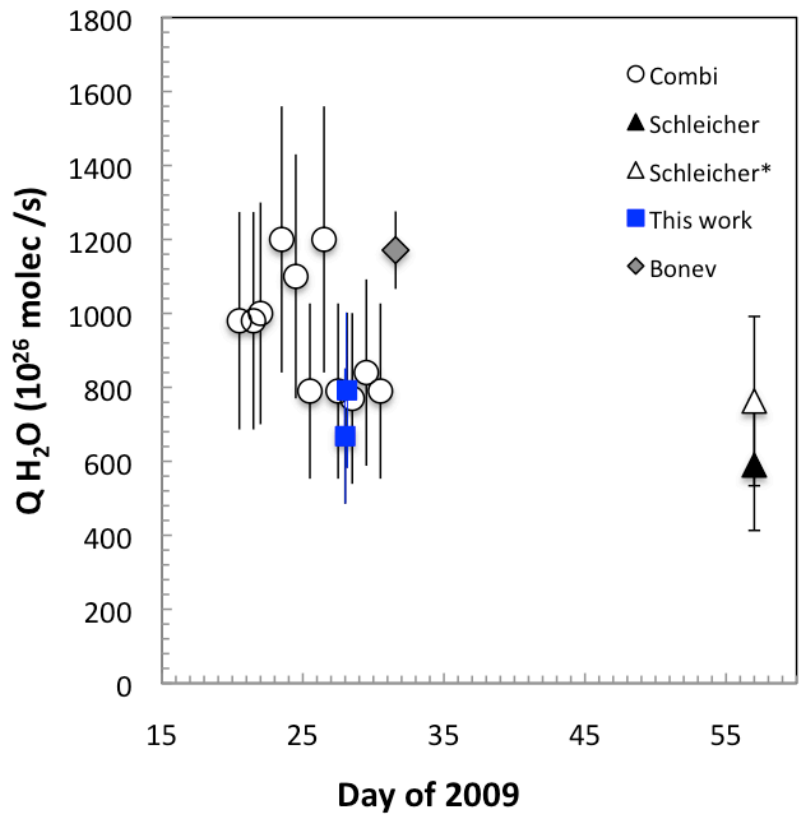

**Figure 7.** Water production rate of comet Lulin. Open circles – Combi et al. (2009); Grey diamond – B.P. Bonev (priv. comm.); Filled triangle – D. Schleicher (priv. comm.); Open triangle – same data, corrected for heliocentric distance. Black squares: this work. Error bars include both stochastic and systematic errors.

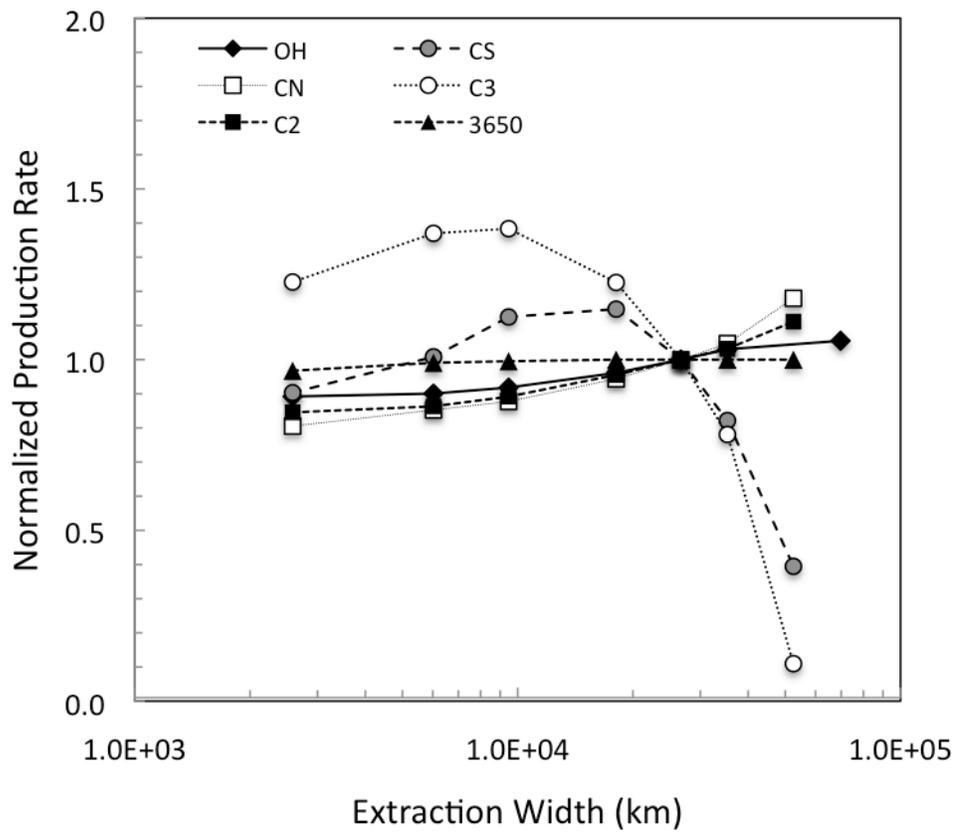

**Figure 8.** Normalized gas production rates with increasing extraction height (h) for the observation at 01:32 UT. The extraction width is the perpendicular distance from the dispersion axis. Data points are normalized to the production rates derived for an extraction height of 27,000 km (15.5 pixels).